\documentclass[a4paper,11pt]{article}
\usepackage{jinstpub} 

\newcommand{\etal}{\rm et~al.}

\usepackage{color}

\title{Semiautomatic dimensional screening of plastic scintillator cubes using image analysis and robotics}

\author[a,1]{Tatsuya Kikawa,\note{Corresponding author.}}
\author[a,2]{Mao Tani,\note{Currently in industry.}}
\author[b]{Atsuko K. Ichikawa,}
\author[c]{Tsunayuki Matsubara,}
\author[a]{Tsuyoshi Nakaya,}
\author[c,2]{Tomohisa Ogawa}

\affiliation[a]{Department of Physics, Kyoto University, Kyoto, Kyoto 606-8502, Japan}
\affiliation[b]{Department of Physics, Tohoku University, Sendai, Miyagi 980-8578, Japan}
\affiliation[c]{High Energy Accelerator Research Organization (KEK), Tsukuba, Ibaraki 305-0801, Japan}

\emailAdd{kikawa.tatsuya.6e@kyoto-u.ac.jp}

\abstract{
Large-scale particle physics detectors often contain millions of repeated components, making precise and efficient quality control essential. We have developed a semiautomatic system for dimensional screening of 1 cm$^3$ plastic scintillator cubes for their potential use in future neutrino detectors. The system employs a motorized rotating stage, six high-resolution cameras, and image analysis software to measure cube size, surface protrusions, and the positions of holes for wavelength-shifting fibers used in optical readout. Based on these measurements, each cube is automatically classified as either acceptable or defective.
We constructed and validated a prototype system, achieving a measurement precision of 10 $\mu$m and over 80\% agreement with manual screening. To enable classification of cubes into 48 groups based on hole positions while preserving their orientation, we introduced a 6-axis robotic arm. The completed system achieved a rejection rate of 3.1\%.
Our approach contributes to scalable, precise, and efficient quality control for future large-scale particle physics detectors.
}


\begin{document}
\maketitle
\flushbottom


\section{Introduction}\label{sec_intro}

In modern particle physics experiments, large-scale detectors such as ATLAS\cite{atlas}, CMS\cite{cms}, or Hyper-Kamiokande\cite{hk} are composed of millions of finely manufactured components whose mechanical precision could critically affect detector performance and robustness. As such detectors become increasingly granular and complex, ensuring consistent quality across components has become a major technical challenge. In particular, quality control procedures must be both precise and scalable to process vast quantities of detector elements within reasonable timeframes and with minimal human error.

\begin{figure}[htbp]
    \begin{center}
         \includegraphics[width=0.35\linewidth]{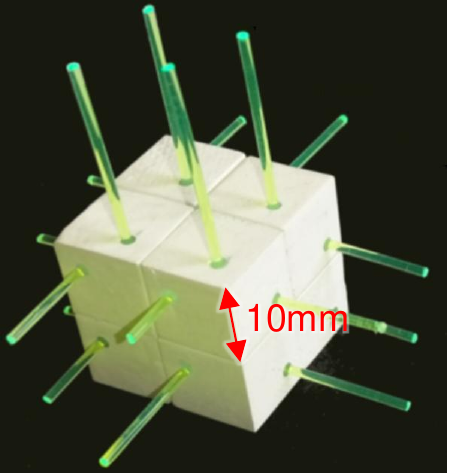}
         \caption{
           2$\times$2$\times$2 scintillator cubes and WLS fibers used for SuperFGD.
         }
         \label{fig:sfgd_cube}
    \end{center}
\end{figure}

This challenge was especially prominent in the upgrade of the near detector in the T2K experiment~\cite{bib:t2k, bib:tdr}, a long-baseline neutrino oscillation experiment in Japan.
A key component of this upgrade is the SuperFGD~\cite{bib:sfgd, bib:sfgd_kikawa, bib:sfgd_paper}, a fully active tracking detector consisting of 1,956,864 plastic scintillator cubes (manufactured by UNIPLAST, Russia), each with a volume of 1~cm$^3$, arranged in a three-dimensional grid.
The sensitive volume of the SuperFGD measures 192~cm $\times$ 182~cm $\times$ 56~cm and has a total mass of around 2 tons. Each scintillator cube (hereafter referred to as a cube) has three orthogonal holes of 1.5~mm diameter, through which 1 mm diameter wavelength-shifting (WLS) fibers (Kuraray Y11(200)M, 1.0~mmD BSJ) are inserted as shown in Fig.~\ref{fig:sfgd_cube}. The scintillation light collected by the fibers is detected at their ends by silicon photomultipliers (SiPMs; Hamamatsu S13360-1325PE).

Each cube was manufactured using injection molding, followed by chemical etching to form a reflective white surface layer of approximately 50–80~$\mu$m, and then drilled with holes from three orthogonal directions. However, variations in the cube dimensions and hole positions inevitably occurred.
When such cubes are densely stacked in a three-dimensional grid, even slight dimensional differences can accumulate and cause alignment issues.

Since the holes are 1.5~mm in diameter and the fibers are 1.0~mm, a misalignment of more than 0.5~mm between adjacent cubes prevents fiber insertion. Along the longest axis of SuperFGD, 192 cubes are aligned, implying that the acceptable deviation per cube must be well below 0.5~mm$ / \sqrt{192} =$ 36 $\mu$m assuming statistically independent deviations of individual cubes that follow a Gaussian distribution. In practice, the standard deviations in cube size and hole position were about 23 $\mu$m and 80 $\mu$m, respectively.
A toy Monte Carlo study indicates that more than half of the fiber holes would become unusable without dimensional quality control.
Moreover, stress on the fibers due to slight lateral misalignments during or after assembly could degrade the optical quality or even break the fibers.

Therefore, quality control was necessary to identify and exclude cubes that deviated significantly from the design.
In Russia, all cubes underwent the following screening procedure prior to assembly~\cite{bib:sfgd_qc, bib:claudio}:
\begin{enumerate}
    \item A total of 15 $\times$ 15 cubes were arranged with aligned hole positions.
    \item Stainless-steel rods of 1.4 mm diameter were inserted into all 15 vertical and 15 horizontal holes (30 rods total). Cubes through which rods could not pass smoothly or which exhibited high insertion stress were identified as defectives and replaced with others until all rods passed smoothly.
    \item The 15 horizontal rods were removed, and each cube was rotated 90 degrees around the vertical rods. Another 15 rods were then inserted from the new horizontal direction, and defective cubes were again replaced.
\end{enumerate}

This screening process resulted in the removal of approximately 5\% of all cubes as defectives.
After the screening of cubes, a plane comprising 192$\times$182 cubes was formed by threading fishing lines through the fiber holes of the cubes.
In total, 56 planes were assembled for the SuperFGD. Then, after the installation of the 56 planes in the SuperFGD mechanical box, the fishing lines were replaced by WLS fibers.
Owing to the cube screening using the stainless-steel rods and the alignment of cube holes using the fishing lines, all fibers were successfully inserted during assembly, with no significant degradation or breakage observed.

However, this screening was performed entirely by hand, requiring considerable human effort and time to process the large number of cubes. Each 15$\times$15 cube batch took over an hour to inspect, corresponding to more than 16 seconds per cube. Furthermore, because the decision to reject a cube was based on the perceived stress on the rods, the process lacked quantitative and reproducible criteria.
In addition, because the screening was performed with cubes assembled in an array, the judgment for an individual cube was not fully independent and could be affected by the quality of neighboring cubes.

Given the excellent performance of the SuperFGD and its potential for future neutrino experiments such as DUNE~\cite{bib:dune}, which is considering the use of larger detectors with similar structures for near detectors, it is essential to develop a more efficient, quantitative, and high-precision quality inspection and screening method.

We have developed and demonstrated a novel quality inspection and screening method that uses image analysis and robotics to evaluate the dimensions of scintillator cubes in an automated and quantitative manner. The remainder of this paper is organized as follows. Section~\ref{sec_system} describes the prototype inspection system based on cameras and a rotating stage. Section~\ref{sec_analysis} explains the image analysis methods. Section~\ref{sec_commissioning} presents the results obtained with the prototype system.
Section~\ref{sec_robot} introduces the completed system incorporating a 6-axis robotic arm along with its performance results.
Finally, Section~\ref{sec_conclusion} concludes the paper.

\section{Scintillator Inspection and Screening System Using Cameras}\label{sec_system}
\subsection{System Overview}

\begin{figure}[htbp]
    \begin{center}
         \includegraphics[width=0.5\linewidth]{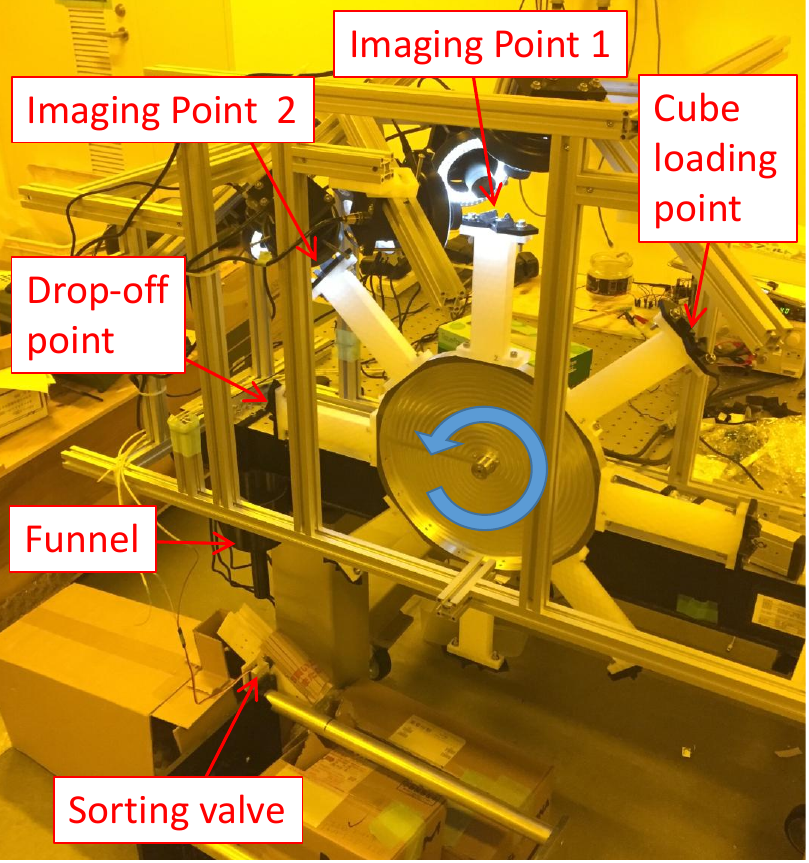}
         \caption{
            Overview of the prototype scintillator inspection and screening system.
         }
         \label{fig:system}
    \end{center}
\end{figure}

The developed scintillator inspection and screening system (hereafter the inspection system) captures images of all six surfaces of a scintillator cube and classifies each cube as either acceptable or defective based on image analysis. An overview of the prototype system is shown in Fig.~\ref{fig:system}.

The cube is placed on a platform, which is mounted on a rotatable stage driven by a stepping motor. Rotating the stage by 45 degrees moves the cube to Imaging Point 1, where three cameras capture images of three surfaces. A further 45-degree rotation brings the cube to Imaging Point 2, where the remaining three surfaces are captured by another set of three cameras.
After the final 45-degree rotation, the cube falls off the platform. Depending on the image analysis result, a sorting valve controlled by a servo motor directs the cube into different boxes.
Since eight platforms are mounted on the stage, different platforms occupy different stations at the same time, allowing loading, imaging, and sorting processes to proceed simultaneously.

The initial placement of the cube on the platform and pressing the trigger key are performed manually. Once the trigger key is pressed, all subsequent steps including image acquisition, analysis, platform movement, and sorting are carried out automatically.
The system evaluates the cubes using optical imaging without applying external mechanical stress during inspection, which also makes the method non-destructive.

\subsection{Optical System}
Six 8-megapixel USB cameras (ELP, ELPUSB8MP02G-SFV(5-50)) are used to image the cubes. Although the resolution is sufficient, the cube surfaces are small (1 cm $\times$ 1 cm), making it difficult to fill the frame while maintaining focus.

To address this, a 2$\times$ teleconverter lens was attached to each camera, enabling full-frame imaging with an effective resolution of approximately 15 $\mu$m per pixel on the cube surface, which meets the requirements for inspection.

\begin{figure}[htbp]
    \begin{center}
         \includegraphics[width=0.45\linewidth]{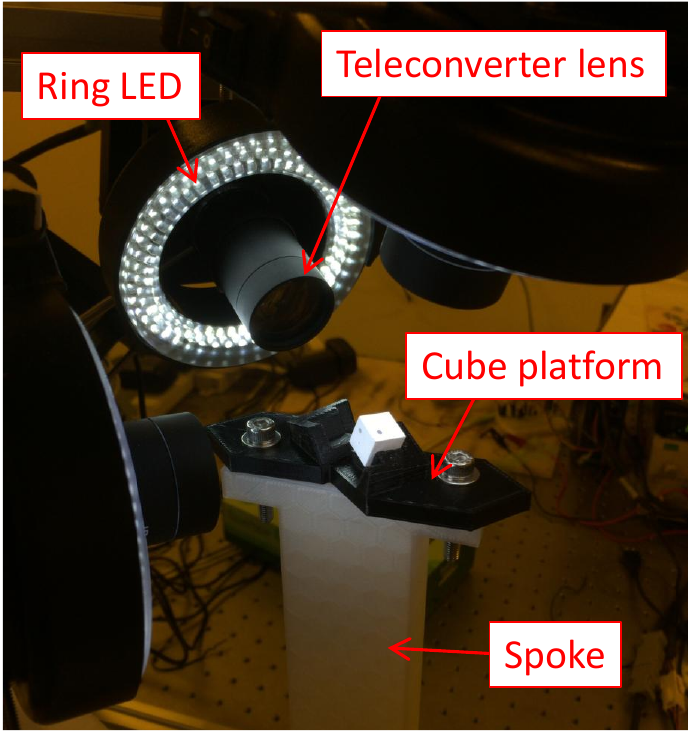}
         \caption{
            Image acquisition unit of the inspection system. 
         }
         \label{fig:camera}
    \end{center}
\end{figure}

Due to surface irregularities introduced by chemical etching, shadows appear when lighting is provided from a single direction, making consistent imaging difficult. To mitigate this, ring-shaped LED lights were mounted on each camera to provide uniform illumination from three directions (Fig.~\ref{fig:camera}).

\subsection{Mechanical System}
The rotation of the stage, which supports the cube platforms via eight spokes, is driven by a 5-phase stepping motor (Oriental Motor, AZM911AC) and its driver (Oriental Motor, AZD-AD). This motor has a positional error of only 0.05 degrees, which does not accumulate over multiple steps.

An octagonal disk with a side length of 120 mm is attached to the motor. Eight spokes extend radially from the disk, each 160 mm long, supporting a cube platform at its end.
The 0.05-degree rotation error translates to a positional deviation of 0.28 mm at the platform, which is sufficiently accurate for imaging purposes.

\begin{figure}[htbp]
    \begin{center}
         \includegraphics[width=0.85\linewidth]{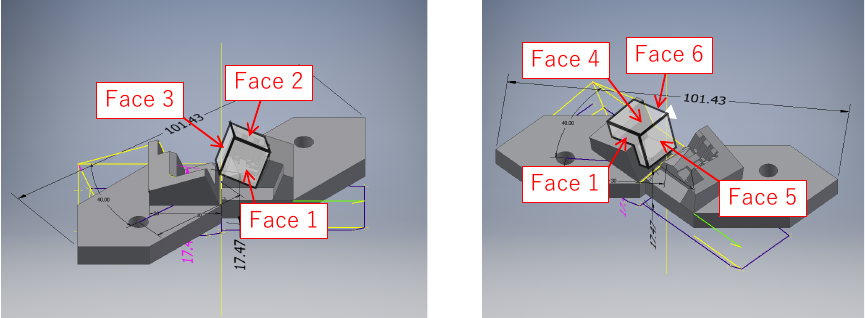}
         \caption{
           Design of the cube platform with a cube in Imaging Point 1 (left) and Imaging Point 2 (right). Labels of opposing faces of the cube are paired as (1,6), (2,5), and (3,4) like a standard die.
         }
         \label{fig:cube_stand}
    \end{center}
\end{figure}

As shown in Fig.~\ref{fig:cube_stand} (left), the cube is initially placed on the right side of the platform. Gravity causes the cube to lean against three walls, resulting in passive and precise positioning. After imaging at Point 1, the cube rolls leftward due to gravity while the platform moves toward Imaging Point 2, revealing the previously hidden three faces for capture, as illustrated in Fig.~\ref{fig:cube_stand} (right).

The platform is 3D-printed using black PLA resin to enhance contrast with the white edges of the cubes. The inner walls are textured to reduce contact area and improve rollability. A fluoropolymer spray is also applied to reduce friction.
The PLA material provides sufficient rigidity for stable cube positioning. Although repeated use could in principle cause gradual wear of the platform edges, no noticeable degradation affecting positioning accuracy was observed during the test runs.

Below the drop-off point, a funnel guides the cube to a sorting valve. A micro servo motor (Tower Pro, SG90) changes the valve’s orientation based on image analysis results, directing the cube into the appropriate box.

All the components, except for the sorting valve and boxes, are fixed to a vibration isolation table either directly or via aluminum frames, ensuring mechanical stability during operation.

\subsection{Control System}
\begin{figure}[htbp]
    \begin{center}
         \includegraphics[width=0.77\linewidth]{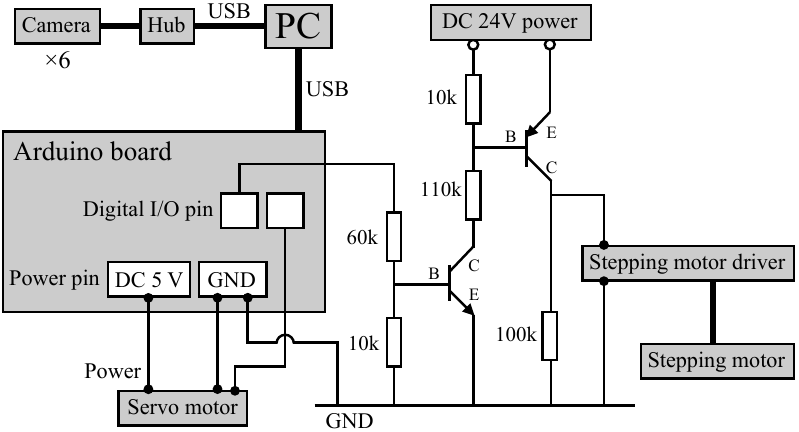}
         \caption{
            Schematic diagram of the control system.
         }
         \label{fig:diagram}
    \end{center}
\end{figure}

Figure~\ref{fig:diagram} shows a schematic overview of the control system.
A single PC controls all six cameras, the stepping motor, and the servo motor.
Once the trigger key is pressed, the PC automatically carries out image acquisition, analysis, platform movement, and sorting in sequence.

All six cameras are connected to the PC via USB through a hub. Image acquisition is handled via the VideoCapture class of the OpenCV library~\cite{bib:opencv, bib:opencvweb}. The image analysis procedure is detailed in Section 3.

An Arduino microcontroller (Arduino Uno R3) is used to interface between the PC and the motors. Arduino provides built-in libraries for the servo control.

The stepping motor driver receives a 24V pulse signal to initiate each movement. Since Arduino outputs only 5V, a custom level-shifting circuit using two transistors was created to convert signals to 24V.

The stepping motor parameters, rotation angle, speed, and acceleration/deceleration profiles, are configured using Oriental Motor’s MEXE02 software. These parameters are tuned to meet specific requirements: smooth rolling on the platform, stable stopping for imaging, and a total inspection cycle of less than 8 seconds per cube.

The software is structured as a loop: it waits for a trigger key, executes the control sequence, and then returns to the wait state.

\section{Image Analysis for Scintillator Inspection}\label{sec_analysis}

In order to determine whether each scintillator cube is acceptable or defective based on its size, shape, and the position and diameter of the holes for WLS fibers, we developed an image analysis method that combines built-in functions from the OpenCV image processing library with custom algorithms.

\subsection{Preprocessing of Images}

First, the captured color images are converted to grayscale. To remove noise caused by pixel defects, dark current, and statistical fluctuations, a Gaussian filter is applied. The images are then binarized using a threshold to classify each pixel as black or white. These steps are essential to ensure that blurry edges due to lighting or surface conditions are not detected as ambiguous or multiple false edges.

\subsection{Cube Contour Detection}

\begin{figure}[htbp]
    \begin{center}
         \includegraphics[width=0.8\linewidth]{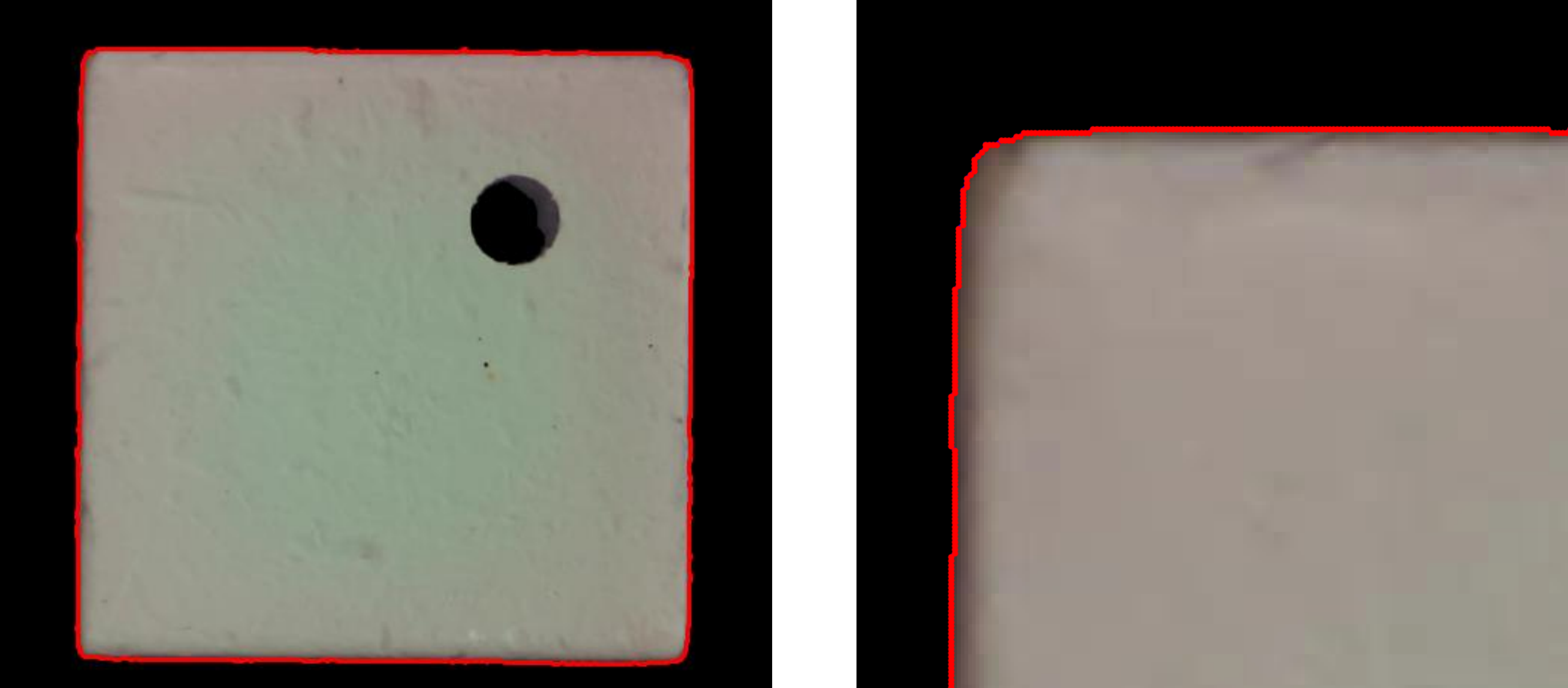}
         \caption{
         Contours detected using OpenCV overlaid on the original image of the full cube (left) and a zoomed-in corner (right).
         }
         \label{fig:cube_edge}
    \end{center}
\end{figure}

Some cubes have small surface protrusions that can interfere with neighboring cubes and must therefore be detected. Using OpenCV, we extract contours of the cube and its holes from the binarized image by detecting the boundaries between white and black pixels (Fig.~\ref{fig:cube_edge}).
The differences between the maximum and minimum coordinates of pixels on the cube contour provide the vertical and horizontal dimensions of the cube, including any protrusions.

\subsection{Shape Detection Using Hough Transform}

\begin{figure}[htbp]
    \begin{center}
         \includegraphics[width=0.8\linewidth]{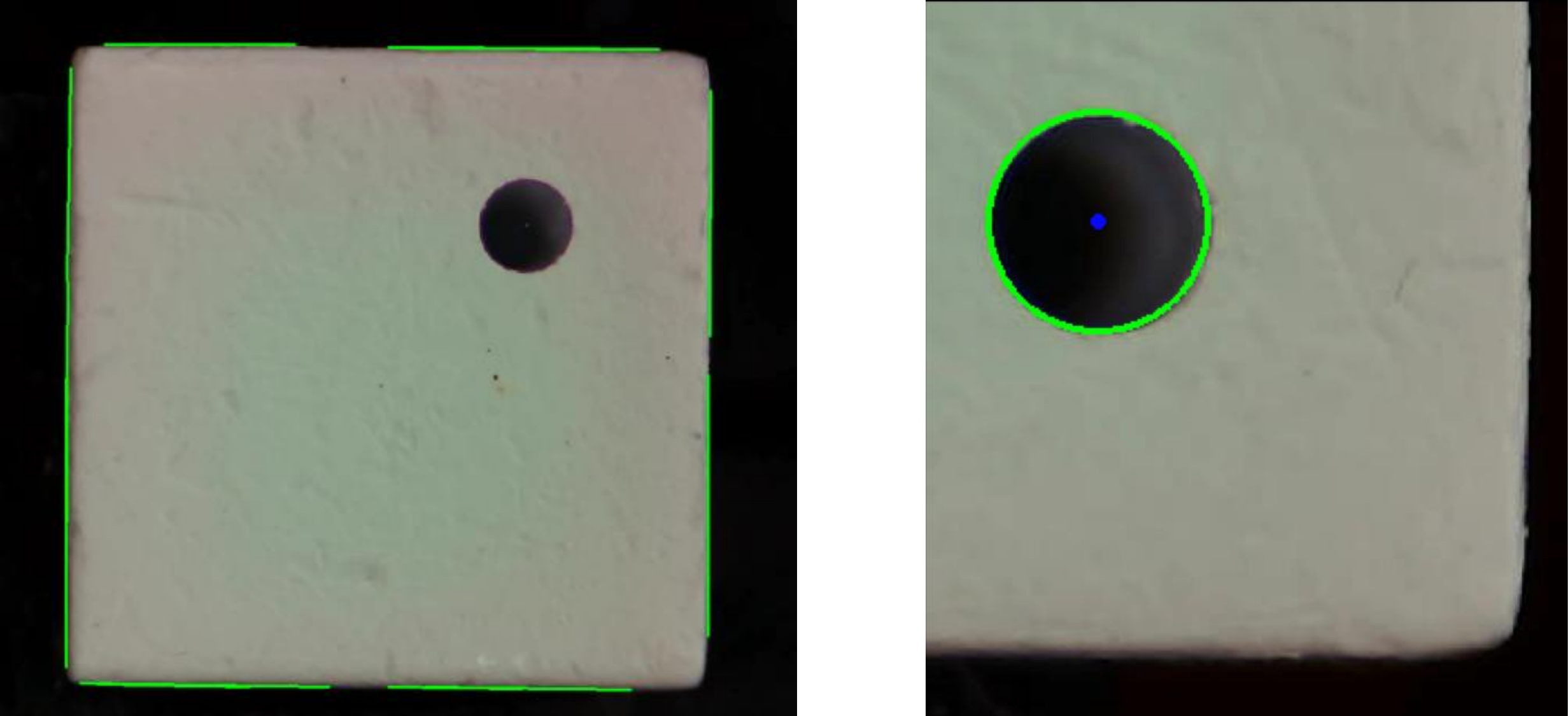}
         \caption{
            Outer edges detected as lines (left) and the hole perimeter as a circle (right) using the Hough transform, overlaid on the original image.
         }
         \label{fig:cube_hough}
    \end{center}
\end{figure}

By applying the Hough transform~\cite{bib:hough} implemented in OpenCV to the contours, we detect the outer edges of the cube as lines and the hole perimeters as circles in the binarized image as illustrated in Fig.~\ref{fig:cube_hough}. Since the Hough transform ignores minor surface irregularities, it provides the cube’s dimensions excluding protrusions, as well as preliminary estimates of the hole's position and size. The hole parameters are later refined through the fitting procedure.

\subsection{Image Alignment Correction}\label{subsec_alignment}

Although the platform passively aligns the cube with high precision and the cameras are carefully positioned, slight systematic tilts can occur and must be corrected. However, the tilt angles derived from the lines detected via the Hough transform are sensitive to surface features and thus not suitable for correction.
Instead, straight lines are fitted to two separated segments along each of the four cube edges, and the tilt is determined from the slope between the centers of these line segments. The average slope from all four cube edges is used to correct the image by rotation. All coordinates of the edges and holes are rotated accordingly, and the contours are reextracted to reflect the corrected geometry.

\subsection{Detection of Surface Protrusions}
To detect protrusions, the extracted contours are compared with the edge lines obtained via the Hough transform. Regions where the contour extends outside the edges are identified as possible protrusions.
Since no special cleaning procedure is applied prior to measurement, small dust particles or fibers may remain on the cube surface and occasionally appear in the captured images. These contaminants may be falsely identified as protrusions.
To avoid such false positives, the fraction of bright pixels in the corresponding area is used as a quantitative measure. Only regions exceeding a specified brightness threshold are classified as actual protrusions.

\subsection{Circular Fitting of Hole Edges}\label{subsec_circular_fit}
The Hough transform alone does not provide sufficient accuracy for hole detection, so further refinement is performed via circle fitting.
First, the region around the hole detected by the Hough transform is cropped to eliminate interference from shadows or surface artifacts. Then, 16 points are extracted from the hole contour at angular intervals of 22.5 degrees around the Hough-derived center.
Finally, these points are used in a least-squares circle-fitting algorithm, initialized with the Hough-derived center and radius, to determine the precise hole position and size.

\subsection{Correction for Lighting and Alignment Effects}\label{subsec_correction}
As described in Section~\ref{sec_system}, each cube face is illuminated from three directions. When a given face is imaged, two of the four side faces surrounding it are illuminated while the other two are not. This asymmetric lighting causes the edges adjacent to the illuminated faces to appear sharper than those adjacent to the unlit faces. This leads to variability in measurement results depending on cube orientation.

Additionally, the system includes eight platforms and six cameras, and slight differences in alignment among them can affect the results. To correct for these effects, a calibration study was performed using 16 cubes, covering all 6 surfaces (96 samples in total). This number of cubes was chosen such that the statistical uncertainties of the correction coefficients were sufficiently small.

\begin{figure}[htbp]
    \begin{center}
         \includegraphics[width=0.95\linewidth]{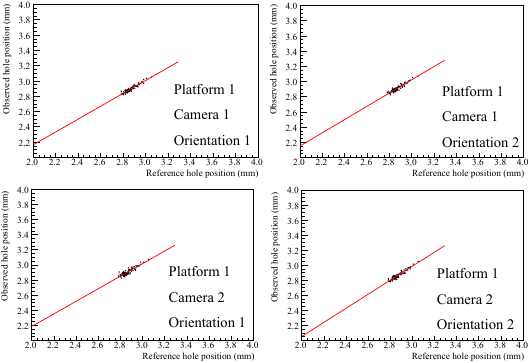}
         \caption{
Example distributions of measured hole positions under various configurations compared with those obtained in the reference configuration.
The red lines indicate the results of the linear fits.
Four out of 192 configurations (6 cameras $\times$ 8 platforms $\times$ 4 orientations) are shown.   
         }
         \label{fig:calibration}
    \end{center}
\end{figure}

For each cube face, 192 images were captured, corresponding to 6 cameras $\times$ 8 platforms $\times$ 4 orientations.
One configuration was arbitrarily selected as the reference from the 192 possible configurations, and the measured hole positions in other configurations were compared with those obtained in the reference configuration, as shown in Fig.~\ref{fig:calibration}.
Here and in the following, the hole positions are defined relative to the adjacent cube edges determined after the alignment correction described in Section~\ref{subsec_alignment}.
Lighting effects appear as an offset (intercept) in the relation, while platform and camera misalignments produce a tilt (slope). A linear fit is therefore used to correct for both effects simultaneously.

\subsection{Cube Classification}\label{subsec_classification}
After the correction, each cube face is evaluated based on five parameters: the hole positions in the horizontal and vertical directions ($x_{hole}, y_{hole}$), the cube size along the horizontal and vertical axes ($L_x, L_y$), and the protrusion index ($P_{bump}$), which quantifies the proportion of bright pixels in regions extending beyond the ideal cube edge.
The classification criteria are as follows:
\begin{itemize}
    \item Defective: If any of the six faces meets one or more of the following:
    \begin{itemize}
        \item $|x_{hole}-2.92$~mm$|>0.21$~mm
        \item $|y_{hole}-2.92$~mm$|>0.21$~mm
        \item $P_{bump}>0.7$
    \end{itemize}
    \item Reinspection required: If not classified as defective, but any face meets:
    \begin{itemize}
        \item $|x_{hole}-2.92$~mm$|>0.19$~mm
        \item $|y_{hole}-2.92$~mm$|>0.19$~mm
        \item $|L_{x}-10.215$~mm$|>0.115$~mm
        \item $|L_{y}-10.215$~mm$|>0.115$~mm
        \item $P_{bump}>0.5$
    \end{itemize}
    \item Acceptable: All remaining cases not meeting the above conditions.
\end{itemize}
These classification criteria were determined empirically from the parameter distributions of cubes manually screened as acceptable or defective using stainless-steel rods. Thresholds were chosen to classify clearly separated regions as “acceptable” or “defective,” while the overlapping region was conservatively assigned to the “reinspection required” category.

\section{Performance of the Prototype Inspection System}\label{sec_commissioning}

We performed a test run of the prototype inspection system using approximately 2,000 scintillator cubes.
A video demonstrating the operation of the inspection system is available in Reference~\cite{bib:mov1}.
The average inspection time per cube was about 6 seconds including the manual cube loading step, which is one third of the time required by the manual method for SuperFGD. In less than 0.5\% of cases, the cube failed to roll properly between Imaging Points 1 and 2, resulting in an incorrect orientation at Point 2. The system classifies such cubes as reinspection targets. Apart from this, the system operated stably without any other malfunction.

\subsection{Evaluation of Measurement Accuracy}
\begin{figure}[htbp]
\begin{center}
    \begin{minipage}{0.47\linewidth}
        \includegraphics[width=0.98\linewidth]{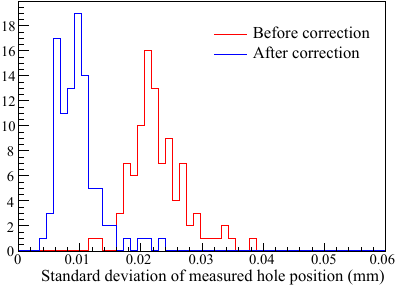}
        \caption{Distribution of the standard deviation of the hole position obtained from multiple measurements. The red and blue histograms correspond to the results without and with correction, respectively.}
        \label{fig:precision}
    \end{minipage}
  \hfill
    \begin{minipage}{0.47\linewidth}
        \includegraphics[width=0.98\linewidth]{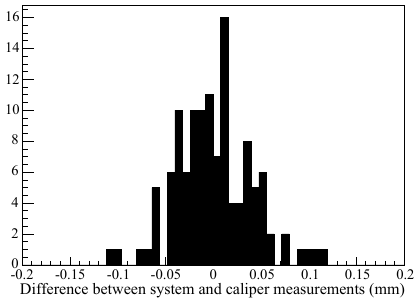}
        \caption{Difference in cube size between the caliper and inspection system measurements.\newline\newline\newline}
        \label{fig:caliper}
    \end{minipage}
\end{center}
\end{figure}

To assess reproducibility, we measured the same cube multiple times with different platforms and orientations and calculated the standard deviation of the results. As shown in Fig.~\ref{fig:precision}, the distribution of standard deviations across 96 cubes shows that the mean deviation was reduced from 23~$\mu$m to 10~$\mu$m by applying the corrections described in Section~\ref{subsec_correction}.
This result indicates that the effects of imaging conditions were effectively mitigated.
The obtained reproducibility is much smaller than the standard deviation of the hole positions among cubes, which is about 80~$\mu$m. These results demonstrate that the measurement performance of the system is well controlled and sufficient for the intended application.

We also compared the inspection system results with caliper measurements for 120 cubes. As shown in Fig.~\ref{fig:caliper}, the standard deviation of the differences between the two methods was 39.8~$\mu$m. This is larger than the inspection system’s reproducibility of 10~$\mu$m, likely due to differences in measurement methodology and caliper measurement errors.

The cube surfaces are slightly uneven because of the chemical etching. When using calipers, the measurement corresponds to the distance between the two specific surface points in contact with the caliper jaws.
To evaluate the reproducibility of the caliper method, repeated measurements were performed on a single cube by varying the contact positions of the caliper jaws within the same pair of opposing surfaces, resulting in a standard deviation exceeding 30~$\mu$m.
In contrast, the inspection system effectively measures the average distance between opposing faces based on the detected straight lines. Additionally, since the surface reflective layer of the cube is soft, caliper measurements may compress the cube slightly, whereas the inspection system performs non-contact measurements.

These factors suggest that the measurements with the inspection system are better suited for quality inspection of the cubes than caliper measurements.

\subsection{Test Run with Manually Screened Cubes}
\begin{figure}[htbp]
    \begin{center}
         \includegraphics[width=0.95\linewidth]{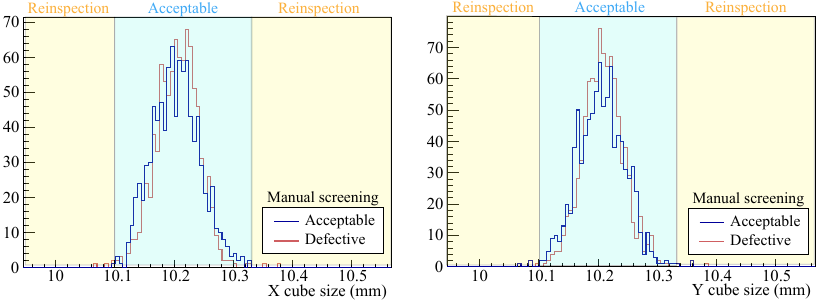}
         \caption{
            Measured cube sizes along the X (left) and Y (right) directions for manually classified samples. The red and blue histograms correspond to cubes manually classified as defective and acceptable, respectively.
         }
         \label{fig:cube_size}
    \end{center}
\end{figure}

\begin{figure}[htbp]
    \begin{center}
         \includegraphics[width=0.95\linewidth]{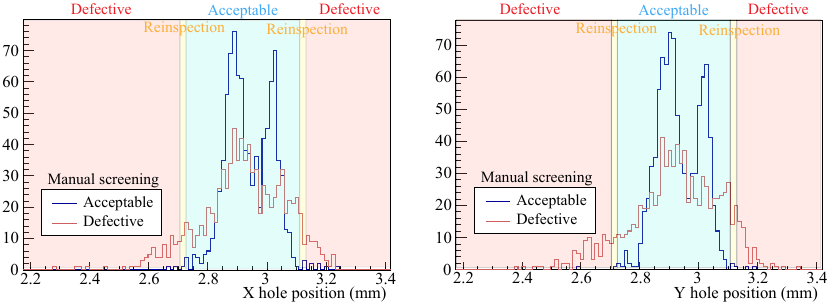}
         \caption{
            Measured hole positions along the X (left) and Y (right) directions for manually classified samples. The red and blue histograms correspond to cubes manually classified as defective and acceptable, respectively. 
         }
         \label{fig:hole_position}
    \end{center}
\end{figure}

We conducted a test using cubes that had previously been manually classified as acceptable or defective using stainless-steel rods. This test sample, consisting of 154 acceptable and 241 defective cubes, is independent from the sample used for optimizing the classification criteria in Section~\ref{subsec_classification}. Cubes flagged for reinspection were reevaluated, and those still classified as reinspection targets were treated as defective. The distributions of cube sizes and hole positions measured by the system are shown in Figs.~\ref{fig:cube_size} and~\ref{fig:hole_position}.
The double-peak structure in the hole-position distributions in Fig.~\ref{fig:hole_position} may be attributed to small machine-to-machine differences among the multiple drilling machines operated in parallel to increase the production speed.

As summarized in Table~\ref{tab:result}, the system showed over 80\% agreement with the manual classification results for both acceptable and defective cubes.
Because manual classification is inherently subjective and operator-dependent, and may also be influenced by neighboring cubes in the tested array, perfect agreement is not expected.

The purpose of this prototype study was to establish a quantitative and reproducible image-based screening scheme and to compare its consistency with the practical reference provided by the manual method. These results suggest that the prototype system can provide a level of practical screening reliability comparable to that of the manual method for a SuperFGD-scale detector.

\begin{table}[htbp]
 \begin{center}
   \caption{Classification results of manually sorted cubes using the inspection system. The values indicate the number of cubes, while percentages in parentheses represent the fraction relative to the total number of manually classified acceptable or defective cubes.}
   \label{tab:result}
\begin{tabular}{lrr} \hline
    & \multicolumn{2}{c}{Manual screening}\\
    & Acceptable & Defective \\ \hline
Acceptable by inspection system & 127 (82.5\%) & 30 (12.4\%) \\
Defective by inspection system  & 27 (17.5\%) & 211 (87.6\%) \\ \hline
\end{tabular}
 \end{center}
\end{table}

\section{Completed System Incorporating a Robotic Arm}\label{sec_robot}

\subsection{Motivation and Overview of Incorporating a Robotic Arm}

Although the prototype inspection system demonstrated high accuracy and reproducibility, it identified approximately 20\% of the cubes as defective under the current criteria. While this defective rate could be reduced by improving the classification criteria, lowering it below 5\% would be difficult.

Furthermore, for the construction of detectors larger than SuperFGD, the accumulation of small misalignments becomes more significant, necessitating even higher precision in alignment. These considerations motivated a reevaluation of the quality control strategy.

The test run described in Section~\ref{sec_commissioning} showed that only about 2\% of the cubes were classified as defective due to size or surface protrusions. The vast majority were rejected based on deviations in hole positions. However, if cubes with similarly displaced holes in the same direction are grouped together in a row, WLS fibers can still be inserted successfully.

To enable such a quality control, it is necessary to classify the cubes into a greater number of categories based on the measured hole positions. However, the existing servo-motor-driven sorting valve only supports classification into at most three categories.
Moreover, since cubes drop from the platform before sorting, the orientation of each cube at the time of imaging is lost.

\begin{figure}[htbp]
    \begin{center}
         \includegraphics[width=0.6\linewidth]{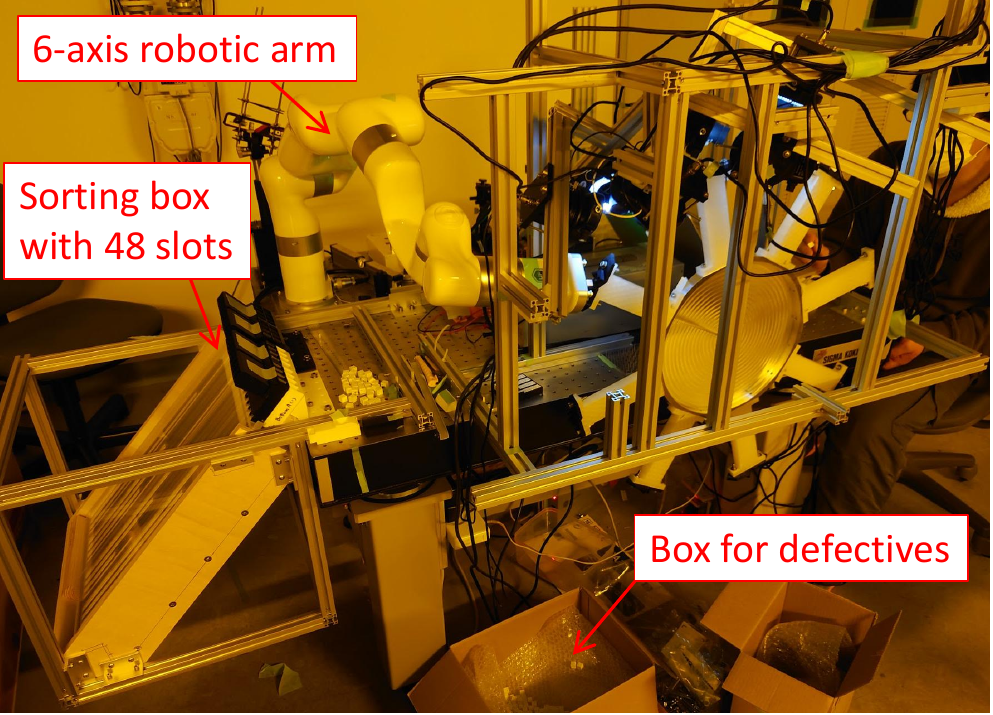}
         \caption{
            Overview of the completed inspection system incorporating the robotic arm.
         }
         \label{fig:robot_arm}
    \end{center}
\end{figure}

To address these issues, a 6-axis robotic arm (UFACTORY xArm 6) was introduced in place of the servo-motor-driven sorting valve, resulting in the completed version of the inspection system as shown in Fig.~\ref{fig:robot_arm}. An optional vacuum gripper was mounted at the tip of the arm, allowing it to pick up cubes by suction and release them by stopping suction. The robotic arm is controllable by a PC via its control box, and we developed custom software integrating the robotic arm with the cameras and stepping motor using the software development kit provided by the manufacturer.

In the prototype system, rotating the platform by 45 degrees after imaging at Imaging Point 2 causes the cube to fall from the platform. However, at Imaging Point 2 the cube is surrounded by cameras on three sides, preventing access by the robotic arm.
To resolve this, the motion was divided into a 22.5-degree rotation, allowing the robotic arm to pick up the cube, and a subsequent 22.5-degree rotation, implemented as a single automated sequence triggered by a key press.

Among the three holes, the one with the smallest deviation from the nominal position is selected, and its corresponding face is designated as the Z-face. The other two faces are labeled X and Y. Based on the measured hole positions on the X and Y faces, the cube is categorized into four groups per face. Combined with the three possible orientations of the Z-face during imaging, the system classifies the cubes into $3\times4\times4=48$ categories.

A custom-designed sorting box contains 48 entry slots corresponding to these categories. The robotic arm places each cube in the appropriate slot, maintaining its orientation from the imaging step. If a cube has size or protrusion defects on any of the X, Y, or Z faces, or if the hole position on the Z face meets the defective criteria defined in Section 3.8, or if the hole positions on the X or Y face significantly deviate from the nominal values, the robotic arm does not pick it up. Such cubes fall into the defective box during the subsequent platform rotation.

The classification scheme and the defective criteria based on the X and Y face hole positions are described in the next subsection, along with the measurement results. The imaging part of the system and the image analysis method remain unchanged from the prototype.

\subsection{Test Operation of the Completed Inspection System}
The completed inspection system was tested using approximately 6,500 scintillator cubes.
A video showing the completed inspection system in operation can be found in Reference~\cite{bib:mov2}.
The addition of the cube pickup step by the robotic arm increased the average inspection time to about 15 seconds per cube including the manual cube loading step. This still surpasses the speed of the manual screening using stainless-steel rods.
Although the cubes occasionally jammed in the sorting box, the robotic arm functioned reliably. There were no significant failures such as the robotic arm failing to grip or dropping the cube.

\begin{figure}[htbp]
    \begin{center}
         \includegraphics[width=0.95\linewidth]{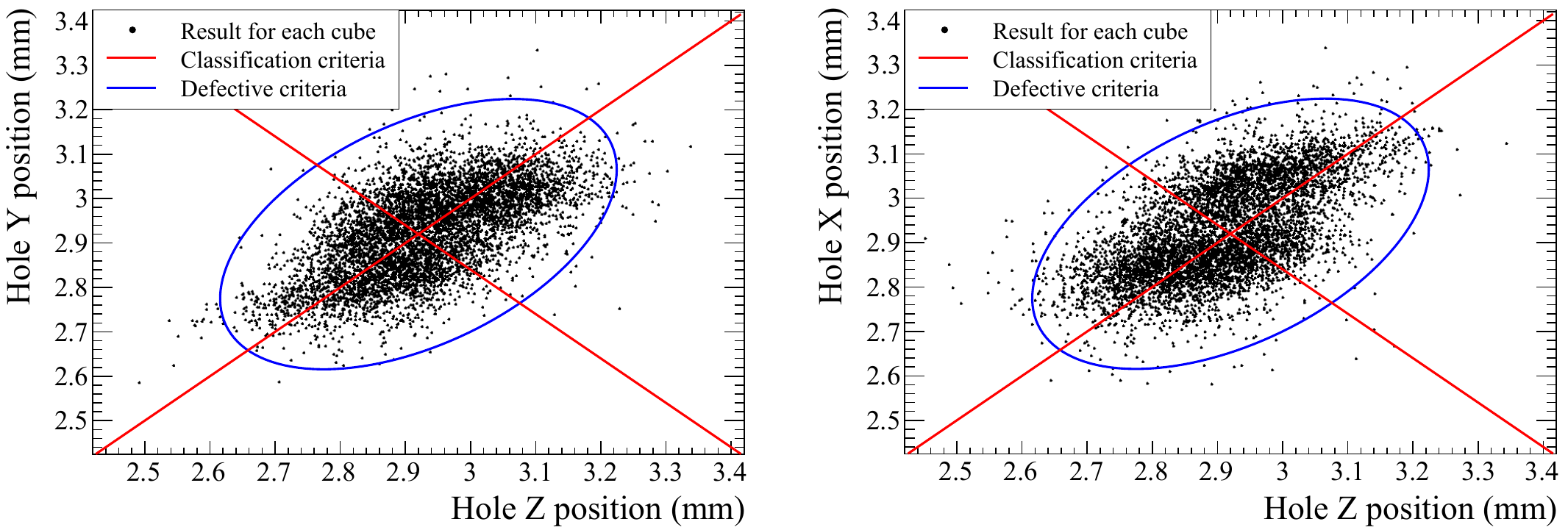}
         \caption{
            Distribution of measured hole positions for the X (left) and Y (right) faces. The red and blue lines indicate the classification and defective criteria, respectively.
         }
         \label{fig:arm_result}
    \end{center}
\end{figure}

The distributions of measured hole positions are shown in Fig.~\ref{fig:arm_result} together with the classification and defective criteria. The classification criteria were determined to distribute the cubes approximately evenly across the groups.
Within each group, the maximum positional variation is below 500 $\mu$m, the allowable tolerance for fiber insertion. This confirms the feasibility of passing a fiber through all holes in a group.
The defective rate was reduced to 3.1\%, representing a significant improvement.

\subsection{Toward Full Automation and Scalable Operation}
The robotic arm is also capable of performing the initial placement of cubes onto the platform which is currently done manually. We demonstrated that the xArm 6 robotic arm can sequentially pick cubes from a densely packed box and place them onto the platform.
Although the current system uses only one robotic arm, introducing a second would enable full automation of the inspection and screening process. Once the box of cubes is set, the system could operate completely unattended.

The 48-category classification increases the complexity of downstream storage and assembly. However, this scheme is intended to be used together with automated handling, in which robotic sorting and systematic category management can reduce the risk of layout errors. In addition, the number of categories can be adjusted according to the required alignment tolerance and the practical constraints of the assembly process.

In large-scale detector construction, the total inspection time is an important consideration. Although the inspection time per cube is approximately 15 seconds in the current system, the overall throughput can be effectively increased by operating multiple inspection units in parallel. Since the system is modular and relatively compact, such parallelization can be implemented straightforwardly, allowing the total processing time to scale inversely with the number of units.

In this study, the calibration was performed once during the initial setup of the system. No significant change in the measurement results was observed during the subsequent inspection of about 6,500 cubes over approximately one month. For larger-scale operation over a longer period, however, periodic recalibration may be required to account for possible environmental changes or mechanical drift.

These features provide a practical path toward fully automated and scalable quality control for future large-scale detector construction.


\section{Conclusion}\label{sec_conclusion}
We developed a semiautomatic quality inspection and screening system for plastic scintillator cubes of the same type as those used in the SuperFGD detector for the T2K experiment. The system measures the cube’s dimensions, surface protrusions, and the positions of holes on three orthogonal faces using a rotating platform, six cameras, and image analysis software. It achieves a measurement precision of 10~$\mu$m and over 80\% agreement with the manual screening results.

However, around 20\% of all cubes were classified as defective in the prototype system. We found that most of these cubes were rejected due to deviations in hole positions. We therefore revised the quality control strategy to utilize these deviations in a constructive way by grouping cubes with similar hole shifts.

To implement this, we introduced a 6-axis robotic arm into the system. The completed system enabled fine-grained classification into 48 categories based on the relative hole positions on each cube face, while preserving their orientation throughout the process. The robotic arm consistently and stably handled the cubes, and the system achieved a rejection rate of 3.1\% in a test run with approximately 6,500 cubes. The classified groups exhibited internal variation in hole positions within acceptable limits for fiber insertion.

Furthermore, we demonstrated that the robotic arm can be used to automate the initial placement of cubes on the platform, paving the way for a fully automated inspection and screening process using two robotic arms.
As particle physics experiments continue to scale up in size and complexity, accurate and efficient quality control of large numbers of components becomes increasingly critical. The results presented in this paper offer a promising approach to addressing this challenge.

\section*{Acknowledgments}
We would like to express our gratitude to the members of the SuperFGD group in the T2K experiment for providing valuable information about the SuperFGD, as well as for offering helpful advice on this research.
In particular, we thank Yury Kudenko for sharing detailed information about the quality control of SuperFGD scintillator cubes conducted in Russia, for facilitating the purchase of scintillator cubes from UNIPLAST for this research, and for providing valuable comments on this paper.
We also appreciate the support of Shunta Arimoto, Masaki Kawaue, Soichiro Kuribayashi, Yohan Lee and Nobuyuki Yoshimura in the test operation of the completed inspection system.
This work was supported by MEXT KAKENHI Grant Number JP16H06288, JP18H05537 and Kyoto University Start-up Grant for Young Researchers, FY2018.



\begin{thebibliography}{99}
  \bibitem{atlas}ATLAS Collaboration, JINST {\bf 3}, S08003 (2008)
  \bibitem{cms}CMS Collaboration, JINST {\bf 3}, S08004 (2008)
  \bibitem{hk}K.~Abe~\etal (Hyper-Kamiokande Proto-Collaboration), arXiv:1805.04163 (2018)
  \bibitem{bib:t2k} K.~Abe~\etal (T2K Collaboration), Nucl. Instrum. Meth. A {\bf 659} 106 (2011)
  \bibitem{bib:tdr} K.~Abe~\etal, arXiv:1901.03750 (2019)
  \bibitem{bib:sfgd} A.~Blondel~\etal, JINST {\bf 13} P02006 (2018)
  \bibitem{bib:sfgd_kikawa} T.~Kikawa~\etal, Nucl. Instrum. Meth. A {\bf 1080}, 170616 (2025)
  \bibitem{bib:sfgd_paper} S.~Abe~\etal, arXiv:2603.14921 (2026)
  \bibitem{bib:sfgd_qc} A. Dergacheva~\etal, Nucl. Instrum. Meth. A {\bf 1041} 11 (2022)
  \bibitem{bib:claudio} C. Giganti~\etal, CERN-SPSC-2020-008, SPSC-SR-267 (2020)
  \bibitem{bib:dune} B. Abi~\etal (DUNE Collaboration), arXiv:2002.02967 (2020)  
  \bibitem{bib:opencv}G. Bradski. Dr. Dobb's Journal of Software Tools (2000)
  \bibitem{bib:opencvweb}OpenCV, Open Source Computer Vision Library, https://opencv.org
  \bibitem{bib:hough}D.H. Ballard, Pattern Recognition 13, 2 (1981)
  \bibitem{bib:mov1}T.~Kikawa, "Operation of the prototype inspection system for plastic scintillator cubes" Zenodo, https://doi.org/10.5281/zenodo.18885041 (2026)
  \bibitem{bib:mov2}T.~Kikawa, "Operation of the completed inspection system incorporating a robotic arm" Zenodo, https://doi.org/10.5281/zenodo.19108914 (2026)
\end{thebibliography}
\end{document}